\pdfoutput=1
\documentclass[usenatbib]{mn2e}
\usepackage{apjfonts}
\usepackage{amssymb}
\usepackage{amsmath}
\usepackage{ctable}
\usepackage{fixltx2e} %% forces figure order to remain fixed (otherwise figure*'s can move out-of-order)

\newcommand{\be}{\begin{equation}}
\newcommand{\ee}{\end{equation}}

\newcommand{\msun}{M_{\sun}}

\newcommand{\paperone}{Paper {\small I}}
\newcommand{\papertwo}{Paper {\small II}}

\newcommand{\flast}{f_{\ell}}
\newcommand{\Sprime}{S^{\prime}}

\newcommand\plotonesize[2]
 {\centering \leavevmode \includegraphics[width={#2\columnwidth}]{#1}}

\newcommand{\acknowledgments}{\begin{small}\section*{Acknowledgments}\end{small}}
\newcommand\altaffilmark[1]{$^{#1}$}
\newcommand\altaffiltext[1]{$^{#1}$}
\title[Stellar Correlation Functions]{Why Do Stars Form In Clusters? An Analytic 
Model for Stellar Correlation Functions\vspace{-0.5cm}}

\author[Hopkins]{
\parbox[t]{\textwidth}{ 
Philip F. Hopkins\altaffilmark{1}\thanks{E-mail:phopkins@astro.berkeley.edu}} 
\vspace*{6pt} \\
\altaffiltext{1}{Department of Astronomy, University of California
  Berkeley, Berkeley, CA 94720\vspace{-1.1cm}} \\
}

\date{Submitted to MNRAS, February, 2012\vspace{-0.6cm}}
\begin{document}
\maketitle
\label{firstpage}

\begin{abstract}

Recently, we have shown that if the ISM is governed by super-sonic turbulent flows, the excursion-set formalism can be used to calculate the statistics of self-gravitating objects over a wide range of scales. On the largest self-gravitating scales (``first crossing''), these correspond to GMCs, and on the smallest non-fragmenting self-gravitating scales (``last crossing''), to protostellar cores. Here, we extend this formalism to rigorously calculate the auto and cross-correlation functions of cores (and by extension, young stars) as a function of spatial separation and mass, in analogy to the cosmological calculation of halo clustering. We show that this generically predicts that star formation is very strongly clustered on small scales: stars form in clustered regions, themselves inside GMCs. Outside the binary-star regime, the projected correlation function declines as a weak power-law, until a characteristic scale which corresponds to the characteristic mass scale of GMCs. On much larger scales the clustering declines such that star formation is not strongly biased on galactic scales, relative to the actual dense gas distribution. The precise correlation function shape depends on properties of the turbulent spectrum, but its qualitative behavior is quite general. The predictions agree well with observations of young star and core autocorrelation functions over $\sim4$\,dex in radius. Clustered star formation is a generic consequence of supersonic turbulence if most of the power in the velocity field, hence the contribution to density fluctuations, comes from large scales $\sim h$. The distribution of self-gravitating masses near the sonic length is then imprinted by fluctuations on larger scales. We similarly show that the fraction of stars formed in ``isolated'' modes should be small, $\lesssim 10\%$.

\end{abstract}

\begin{keywords}
star formation: general --- galaxies: formation --- galaxies: evolution --- galaxies: active --- 
cosmology: theory
\vspace{-1.0cm}
\end{keywords}

\vspace{-1.1cm}
\section{Introduction}
\label{sec:intro}

A central tenet of our understanding of star formation is that star formation is clustered or correlated \citep[e.g.][and references therein]{lada:2003.embedded.cluster.review,
portegies-zwart:2010.starcluster.review}. Observational evidence for this comes from a number of channels: 
observations have directly measured 
the correlation between young O-stars and other populations \citep{oey:2004.stellar.clustering,parker:2007.ob.corr.fn}, and shown that 
most can be directly identified as part of young clusters/associations \citep{gies:1987.ob.associations}, 
with most of the remainder being identifiable as runaways \citep{de-wit:2005.ostar.runaways,schilbach:2008.ostar.runaways}. 
The observed star formation rate in embedded clusters in the MW can account for 
most of the large-scale average SFR \citep{lada:2003.embedded.cluster.review}. 
The correlation function of stars 
in various MW star-forming regions has been measured and rises continuously on 
small scales, over $>4-5$ orders of magnitude in radius \citep{gomez:1993.stellar.corrfn,
larson:1994.hierarchical.sf,simon:1997.stellar.clustering,nakajima:1998.stellar.clustering,
clarke:2000.starcluster.formation.orion,hartmann:2002.taurus.stellar.clustering,
hennekemper:2008.smc.stellar.clustering,kraus:2008.stellar.clustering}. 
In nearby galaxies the observations are more difficult, but 
in starburst galaxies a large fraction of all the UV light is often identified with just a 
few young star clusters, themselves clustered \citep[see e.g.][and references therein]{zhang:2001.antennae.starcluster.clustering}. 

Theoretically, this is broadly understood as a consequence of dense gas being 
concentrated in primarily in GMCs, which then undergo fragmentation and turn 
some fraction of their mass into stars. And it does appear the same clustering 
is evident in the pre/proto-stellar core population \citep{stanke:2006.core.mf.clustering}.
However, this does not provide a 
quantitative model for their correlation function, nor does it actually explain why this should be the case -- why do most stars not form in relatively more isolated high-density regions? 

Although much theoretical progress has been made in understanding how stars form in clusters, our fundamental understanding of why their formation is clustered remains quite limited. 
There is no analytic theory for the star-star correlation function, outside of very small scales where it is dominated by the binary star population. Much of the discussion in the literature has focused on determining the fractal dimension of star formation: if the ISM were structured in a hierarchically fractal manner, this would imply a simple power-law correlation function from which the fractal dimension can be determined. 
But this is not a {physical}, predictive model; it is a parameterization -- albeit a useful one -- of the observations \citep[see also][]{bate:1998.stellar.corrfn.interp}. Moreover it must break down at both small scales (the binary regime) and large scales, since it is also observationally established that star formation is not strongly clustered/biased relative to the dense gas in the disk on very large scales \citep{zhang:2001.antennae.starcluster.clustering,leroy:2008.sfe.vs.gal.prop,foyle:2010.spiral.arms.conc.gas.not.sf}. 

At first glance, it is not surprising that there is no more sophisticated 
analytic theory for the correlation function of stars. The process is highly non-linear, chaotic, and involves a wide range of physics including turbulence, cooling, magnetic fields, and feedback, and the correlation function is a spatially dependent quantity. There have therefore been some attempts to compare stellar correlation functions with numerical simulations of star formation in clusters \citep[see][]{klessen:2000.cluster.formation,hansen:2012.lowmass.sf.radsims}, but the dynamic range of the problem makes this very difficult. Galaxy-scale simulations are required to follow the formation of GMCs and predict their global properties, and how these overdense regions compare to the density field in more isolated regions (i.e.\ why there are not more stars in non-clustered regions). But these cannot resolve the formation of stars and model it with sub-grid recipes -- the typical ``star particle'' in such simulations is the mass of a real star cluster. Simulations which have the necessary resolution can only follow small ``protocluster'' regions whose properties must be assumed as initial conditions; they can predict quantities such as the binary distribution, but not global clustering.

However, \citet{hopkins:excursion.ism} (hereafter \paperone) recently generalized the excursion set formalism -- well known from cosmological applications as a means to calculate halo mass functions and clustering -- to calculate the statistics of bound objects in the density field of the turbulent ISM. The key property of supersonic turbulence that makes this possible is that the density distribution (at least outside of collapsed regions) tends towards a lognormal, with a dispersion that varies in a well-defined manner with Mach number \citep[see e.g.][]{vazquez-semadeni:1994.turb.density.pdf,padoan:1997.density.pdf,ostriker:1999.density.pdf}. In \paperone, we use this to construct the statistics of the ``first-crossing distribution'': the statistics of bound objects defined on the largest scales on which they are self-gravitating. We showed that the predicted mass function and correlation functions/clustering properties agree well with observations of GMCs on galactic scales. In \citet{hopkins:excursion.imf} (\papertwo), we extended the formalism to the ``last crossing distribution'' -- specifically, the mass function of bound objects defined on the smallest scales on which they remain self-gravitating but do not have self-gravitating sub-regions (i.e.\ are not fragmenting). We argued that these should be associated with proto-stellar cores, and in \papertwo\ showed that the resulting core MF agrees well with canonical MW CMF and (by extrapolation) stellar IMFs. This formalized the approximate argument for the same in \citet{hennebelle:2008.imf.presschechter}, but more importantly for our purposes here, places it in the proper excursion set formalism and so makes it possible to calculate higher-order statistics such as clustering properties, using only information on global (galactic) scales.

In this Letter, we use the excursion set formalism, in analogy to its well-studied application for the clustering of dark matter halos, to develop a fully analytic theory for the clustering of protostellar cores and, by extension, star formation, in a turbulent ISM. 

Before going forward, we must make a critical distinction between ``clustered'' star formation and star formation ``in (bound) clusters.'' Whether or not a population is ``clustered'' is a general statement about whether it follows a non-zero two-point correlation function (i.e.\ if members of the population are more likely to appear near other members of the population, relative to their spatial distribution if the objects were randomly distributed in the volume). This is the sense we will refer to throughout the paper, for which we can calculate the correlation function and clustering amplitude, for any scale on which the clustering is evaluated. Whether stars are ``in clusters'' depends on the definition of ``a cluster,'' which is not generic; long-lived clusters must be gravitationally bound, which requires consideration of quantities like the mass expelled as stars form and evolve (not included in our study here). This is a very different question, and is outside the scope of this paper \citep[see e.g.][and references therein]{bressert:2010.clustered.sf.vs.criteria}. 

In \S~\ref{sec:twobarrier} we derive the solution for the statistics of the last-crossing distribution inside a large-scale over/underdensity (the mathematical underpinning of the correlation function). 
In \S~\ref{sec:massfn} we show how this relates to the conditional mass function of cores. 
In \S~\ref{sec:corrfn} we use this to derive the star-star correlation functions. 
In \S~\ref{sec:results} we present the calculated correlation functions as a function of stellar mass and spatial scale, as well as global turbulent ISM properties, and compare to observations. 
In \S~\ref{sec:discussion} we summarize our results and discuss their implications.

\begin{figure}
    \centering
    \plotonesize{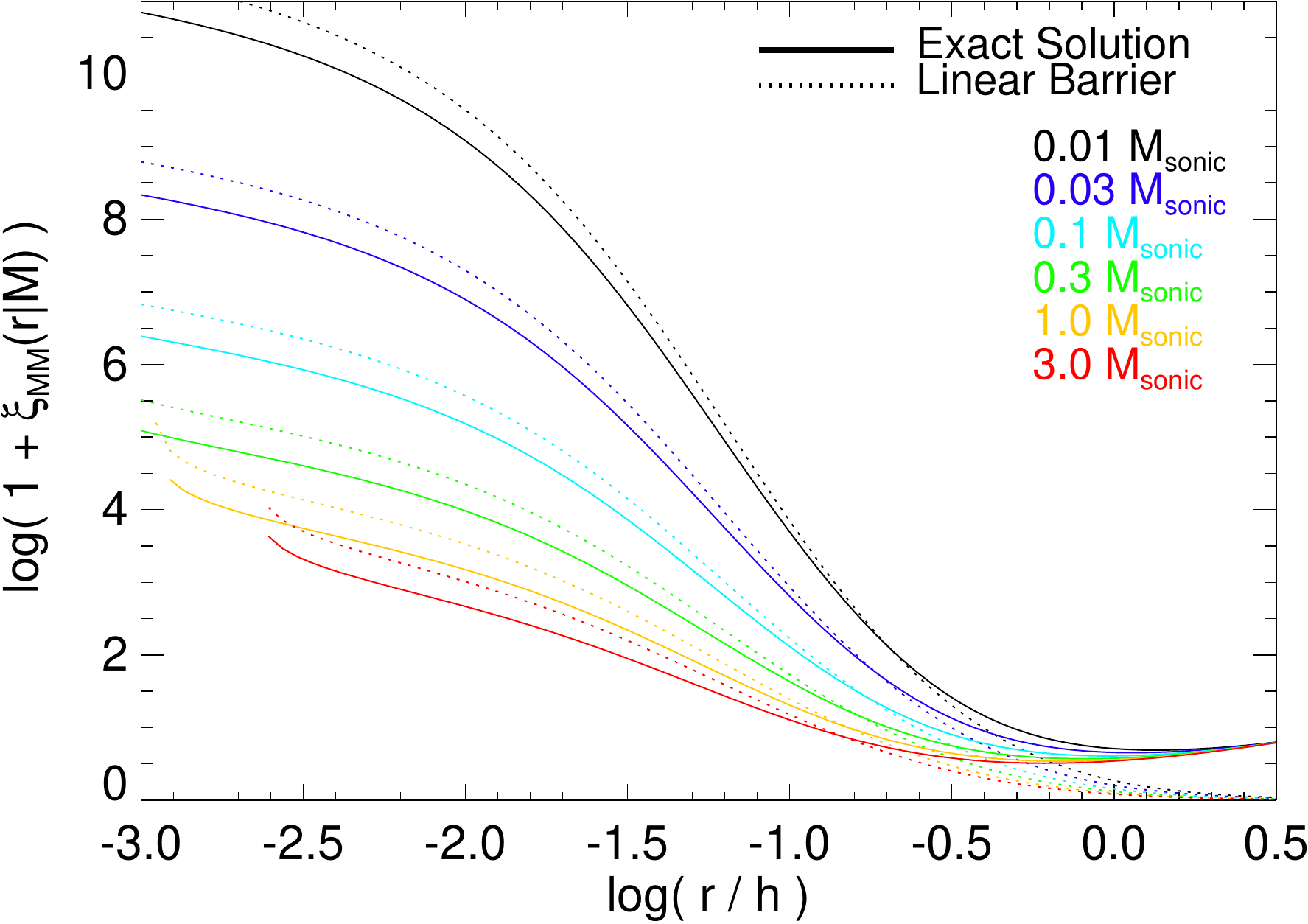}{0.95}
    \caption{The predicted 3-dimensional auto-correlation function 
    of collapsing cores, as a function of radius (in units of the disk scale height $h$).
    The excursion-set model for collapsing cores is determined by solving the 
    two-barrier problem for bound/collapsing objects defined at ``last crossing'' -- i.e.\ 
    the smallest scales on which they are self-gravitating. The model is completely specified 
    for a given turbulent spectral index $p$ and normalization (which we 
    take to be the Mach number at the scale $h$, $\mathcal{M}_{h}$). Here we 
    take $p=2$, $\mathcal{M}_{h}=10$.
    Solid lines show the exact solution from Eq.~\ref{eqn:autocorr}. 
    Dotted lines compare the approximate scaling in Eq.~\ref{eqn:autocorr.approx} -- 
    the normalization is systematically too large, but the shape is correct up to scales 
    near $\sim h$, where the assumption that the run in $S(R)$ is small breaks down.
    The shape is similar at all stellar masses, but the small-scale amplitude increases 
    strongly at low masses. In all cases, cores/stars are predicted to cluster very strongly 
    on small scales, below their parent GMC scale. On large scales ($\gtrsim h$), they 
    are only weakly clustered. The characteristic scale is set by the fact that most of the 
    turbulent velocity power (hence power in density fluctuations) is at this scale.
    \label{fig:autocorr}}
\end{figure}

\begin{figure}
    \centering
    \plotonesize{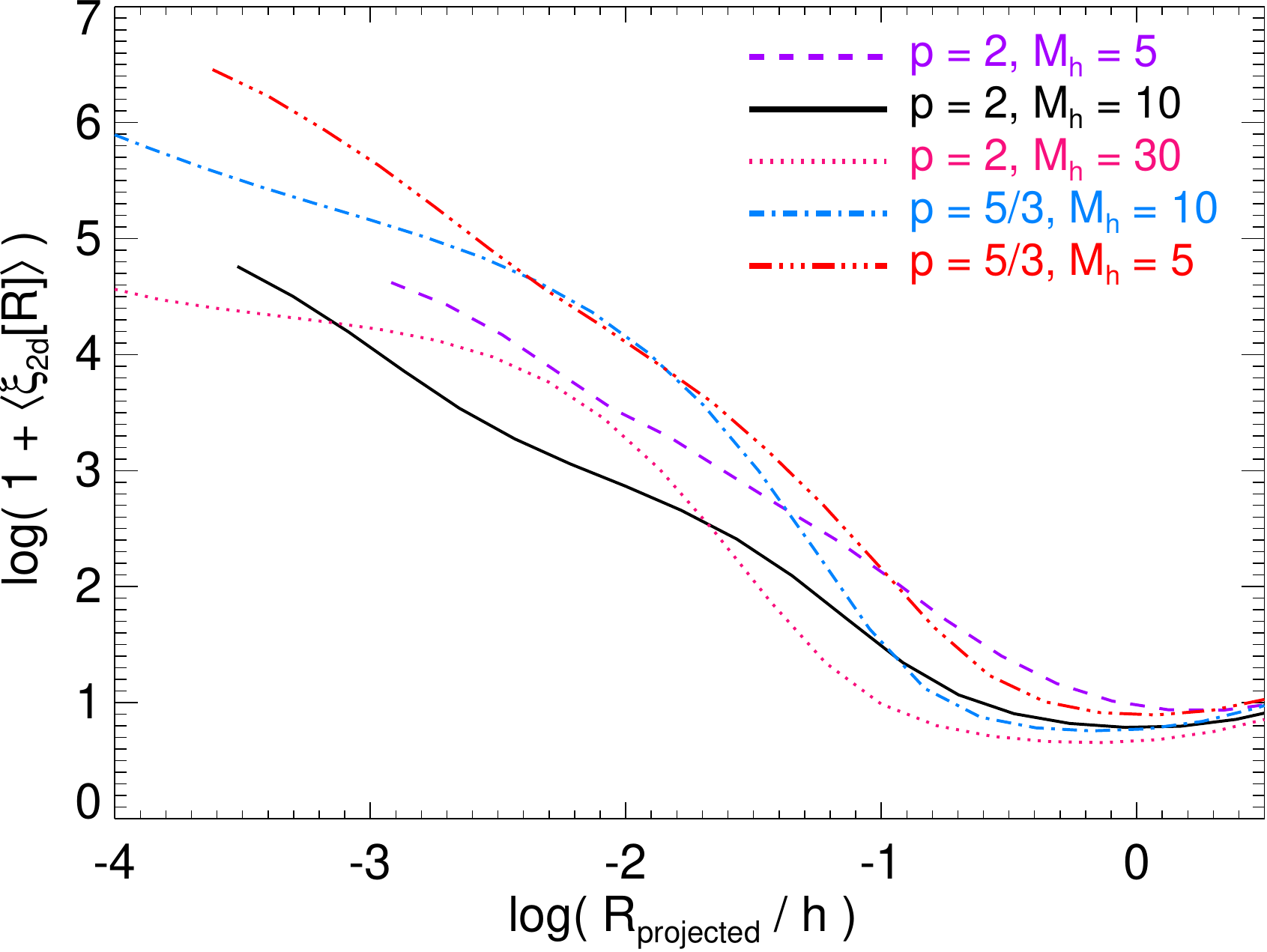}{0.95}
    \caption{The (exact) predicted 2-dimensional correlation function between 
    cores of all masses -- this should be comparable to the observed (projected) 
    star-star correlation function (averaged over the IMF range of stellar masses). 
    The predictions are qualitatively similar in all cases, but complicated 
    second-order differences arise.
    Models with turbulent spectral index 
    $p=5/3$ are similar to those with $p=2$ but a higher $\mathcal{M}_{h}$ 
    by a factor $\sim2$, $\mathcal{M}(r)\propto r^{(p-1)/2}$ falls off more slowly at the 
    small scales of interest. 
    At fixed $p$, lower values of $\mathcal{M}_{h}$ produce slightly steeper, 
    more power-law like behavior. The more pronounced flattening of 
    $\xi$ on small scales with higher $\mathcal{M}_{h}$ 
    arises because the line-of-sight integral is dominated by the (larger) power 
    near large scales. 
    \label{fig:corr.vs.param}}
\end{figure}

\vspace{-0.0cm}
\section{The Two-Barrier Last-Crossing Problem}
\label{sec:twobarrier}

Calculating clustering properties in the excursion set formalism depends on 
the solution to the ``two-barrier problem'' -- the probability of an ``event'' given 
a specific value of the Gaussian field at some larger scale. 
This is well-studied for the first-crossing events, but has not been 
calculated for the last-crossing distribution; we therefore calculate this here. 
Our derivation closely follows that in \papertwo.

Consider the Gaussian field $\delta({\bf x}\,|\,R)$ (which here represents the 
logarithmic density field smoothed in a kernel of radius $R$ about ${\bf x}$). 
The variance $S(R)$ is 
monotonic so we take $S$ as the independent variable and consider 
$\delta(S)$ (for convenience, we will drop explicit notation of ${\bf x}$). 
The PDF of $\delta(S)$ is, by definition, 
\be
P_{0}(\delta\,|\,S) = \frac{1}{\sqrt{2\pi\,S}}\,\exp{\left(-\frac{\delta^{2}}{2\,S} \right)}
\ee

The barrier $B(S)$ is the minimum value $\delta(S)$ which defines 
objects of interest (e.g.\ collapsing regions). 
Consider a ``trajectory'' $\delta(S)$ which begins at some $\delta_{i}$ 
at $S_{i}(R_{i}\rightarrow 0)$, then evaluate it at successively larger scales 
(smaller $S$). We define last-crossing distribution in \papertwo, 
$\flast(S\,|\,\delta_{i})\,{\rm d}S$, as the probability that the trajectory crosses $B(S)$ for the first time 
between $S$ and $S+{\rm d}S$ {\em without} having crossed $B(S)$ at any larger $S$. 
The sum over all trajectories is just given by 
$\flast(S)\equiv\int \flast(S\,|\,\delta_{i})\,P(\delta_{i})\,{\rm d}\delta_{i}$. 

Define $\flast(S\,|\,\delta_{0})$ as the value of $\flast(S)$, for trajectories which 
have a value $\delta(S_{0})=\delta_{0}$ at some {\em larger} scale $S_{0}$.
Also define $\Pi(\delta[S]\,|\,\delta_{0}[S_{0}])\,{\rm d}\delta$ as 
the probability for a trajectory -- constrained to have $\delta_{0}(S_{0})$ -- to have a value between 
$\delta$ and $\delta+{\rm d}\delta$ at scale $S$, {\em without} having crossed $B(S)$ at 
any larger $S^{\prime}>S$. 
The integrals of $\flast(S\,|\,\delta_{0})$ and 
$\Pi(\delta[S]\,|\,\delta_{0}[S_{0}])$ for $\delta<B(S)$ 
must sum to unity: 
\be
\label{eqn:integral.eqn}
1 = \int_{S}^{S_{i}}{\rm d}\Sprime\,\flast(\Sprime\,|\,\delta_{0}) + 
\int_{-\infty}^{B(S)}\Pi(\delta[S]\,|\,\delta_{0}[S_{0}])\,{\rm d}\delta
\ee
If we ignored the barrier, $\Pi(\delta[S]\,|\,\delta_{0}[S_{0}])$ 
would be equal to $P_{10}(\delta[S]\,|\,\delta_{0}[S_{0}])$, 
the probability that a trajectory with $\delta_{0}(S_{0})$ 
has a value $\delta(S)$ at a larger $S>S_{0}$. 
This is just 
\be
P_{10}(\delta[S]\,|\,\delta_{0}[S_{0}]) = P_{0}(\delta-\delta_{0}\,|\,S-S_{0})
\ee
But we must subtract from this the probability that a trajectory crosses 
the barrier at some larger $\Sprime>S$ and then passes through 
$\delta(S)$, so 
\begin{align}
\label{eqn:Pidef}
 \Pi(\delta[S]\,|\,\delta_{0}[S_{0}]) &= 
P_{10}(\delta[S]\,|\,\delta_{0}[S_{0}]) -  \\
\nonumber&\int_{S}^{S_{i}}{\rm d}\Sprime\,\flast(\Sprime\,|\,\delta_{0})\,
P_{01}(\delta[S]\,|\,\delta^{\prime}=B[\Sprime],\,\delta_{0}[S_{0}])
\end{align}
Where $P_{01}(\delta[S]\,|\,\delta^{\prime}[\Sprime],\,\delta_{0}[S_{0}])$ is the 
probability of a transition from 
$\delta(\Sprime)$ to $\delta(S)$ where $\Sprime>S$ -- i.e.\ moving 
in the ``opposite'' direction (decreasing $S$) from the transition 
which defined $P_{10}$ -- for a trajectory subject to the 
constraint that it must equal $\delta_{0}$ at $S_{0}$. 
This can be determined from Bayes's theorem:
\be
\label{eqn:P01}
P_{01}(\delta[S]\,|\,B[\Sprime]) = 
P_{10}(B[\Sprime]\,|\,\delta[S])\,
\frac{P_{10}(\delta[S]\,|\,\delta_{0}[S_{0}])}
{P_{10}(B[\Sprime]\,|\,\delta_{0}[S_{0}])}
\ee
where on the right-hand side we have $P_{10}(\delta[S]\,|\,\delta_{0}[S_{0}])$ 
and $P_{10}(B[\Sprime]\,|\,\delta_{0}[S_{0}])$ instead of $P_{0}(\delta\,|\,S)$ 
and $P_{0}(B[\Sprime]\,|\,\Sprime)$ because the distributions are constrained 
to have $\delta_{0}(S_{0})$. But because fluctuations on successive scales 
are uncorrelated, the probability $P_{10}(B[\Sprime]\,|\,\delta[S])$ does not 
explicitly depend on $\delta_{0}$.

The governing equation for $\flast(S\,|\,\delta_{0})$ is now 
completely defined in terms of normal distributions. 
We can therefore follow exactly the procedure in 
\papertwo. After differentiating Eq.~\ref{eqn:integral.eqn} to isolate 
$\flast(S\,|\,\delta_{0})$ and performing a considerable amount of 
simplifying integral evaluation,\footnote{Taking ${d}/{d}S$ of Eq.~\ref{eqn:integral.eqn}, we obtain
\be
\nonumber
\flast(S|\delta_{0}) = \frac{dB}{dS}\,\Pi(B[S]|\delta_{0}[S_{0}]) + \int_{-\infty}^{B(S)}d\delta\,\frac{d}{dS}\,\Pi(\delta[S]|\delta[S_{0}])
\ee
Noting $S^{\prime}>S>S_{0}$, Eq.~\ref{eqn:P01} can be simplified to 
\be
\nonumber
P_{01}(\delta[S]\,|\,B[\Sprime]) = P_{0}(\delta[S]-\delta_{e}[S,\,S^{\prime}]\,|\,S_{e}[S,\,S^{\prime}])
\ee
where $\delta_{e}$ and $S_{e}$ are defined in Eqs.~\ref{eqn:deff.def}-\ref{eqn:Seff.def}.
We then expand $\Pi$ using Eq.~\ref{eqn:Pidef}, take the derivatives with respect to $S$ where appropriate, and use the simplifying relations:
\begin{align}
\nonumber \int_{-\infty}^{B[S]}\,d\delta\,\frac{d}{dS}\,P_{0}(\delta-\delta_{0}|S-S_{0}) &= -\frac{B[S]-\delta_{0}}{2\,(S-S_{0})}\,P_{0}(B[S]-\delta_{0}|S-S_{0}) \\ 
\nonumber {\rm Limit}{\Bigl[}\int_{-\infty}^{B[S]}\,d\delta\,P_{01}&(\delta[S]\,|\,B[S^{\prime}]){\Bigr]}_{S^{\prime}\rightarrow S} = \frac{1}{2} \\ 
\nonumber \int_{-\infty}^{B[S]}{d\delta}\frac{d}{dS}\,P_{01}(\delta[S]\,|\,B[\Sprime]) &= \\
-\frac{1}{2}\,{\Bigl[}
\frac{B(S^{\prime})-B(S)}{S^{\prime}-S} + &\frac{B(S)-\delta_{0}}{S-S_{0}}{\Bigr]}\,
P_{0}(B[S]-\delta_{e}[S,\,S^{\prime}]\,|\,S_{e}[S,\,S^{\prime}])\nonumber
\end{align}
} we arrive at the 
key equation for $\flast(S\,|\,\delta_{0})$: 
\begin{align}
\label{eqn:flast.2barrier}
\flast(S\,|\,\delta_{0}) = g_{1}(S\,|\,\delta_{0}) + 
\int_{S}^{S_{i}}\,{\rm d}S^{\prime}\,\flast(S^{\prime}\,|\,\delta_{0})\,g_{2}(S,\,S^{\prime}\,|\,\delta_{0})
\end{align}
where 
\begin{align}
g_{1}(S\,|\,\delta_{0}) &= {\Bigl [}2\,\frac{dB}{dS} -\frac{B(S)-\delta_{0}}{S-S_{0}}{\Bigr]}\,
P_{0}(B(S)-\delta_{0}\,|\,S-S_{0})\\
g_{2}(S,\,\Sprime\,|\,\delta_{0}) &= {\Bigl[}\frac{B(\Sprime)-B(S)}{\Sprime-S} 
+\frac{B(S)-\delta_{0}}{S-S_{0}}-2\,\frac{dB}{dS}{\Bigr]}\,
\times\\
\nonumber& \ \ \ \ \ \ \ \ \ \ P_{0}{\Bigl[}B(S) - \delta_{e}(S,\,\Sprime)\, |\, S_{e}(S,\,\Sprime) {\Bigr]}\\
\label{eqn:deff.def}\delta_{e}(S,\,\Sprime) &= B(\Sprime) + (\delta_{0}-B[\Sprime])\,{\Bigl(}\frac{\Sprime-S}{\Sprime-S_{0}}{\Bigr)}\\
\label{eqn:Seff.def} S_{e}(S,\,\Sprime) &= (\Sprime - S)\,{\Bigl(}\frac{S-S_{0}}{\Sprime-S_{0}}{\Bigr)}
\end{align}
Although complicated, Eq.~\ref{eqn:flast.2barrier} has, in general, a unique solution for any arbitrary barrier $B(S)$, and is straightforward to solve via standard numerical methods for any choice $\delta_{0}(S_{0})$ provided $S_{0}<S$ (see the discussion in \papertwo\ and \citet{zhang:2006.general.moving.barrier.solution}). 

In fact for a linear barrier, $B(S)=B_{0}+\beta\,S$, this has a remarkably simple closed-form solution: 
\be
\label{eqn:linear}
\flast(S\,|\,\delta_{0})|_{B=B_{0}+\beta\,S} = \beta\,P_{0}(B[S-S_{0}]-\delta_{0}\,|\,S-S_{0})
\ee

It is easy to verify that when $\delta_{0}\rightarrow0$ as $S_{0}\rightarrow0$, we recover the one-barrier last-crossing distribution from \papertwo, which is the value of $\flast(M\,|\,\delta_{0})$ averaged over all 
$\delta_{0}(S_{0})$
\be
\flast(M)\equiv\langle \flast(M\,|\,\delta_{0})\rangle = \flast(M\,|\,\delta_{0}\rightarrow0,\,S_{0}\rightarrow0)
\ee

\vspace{+2.5cm}
\section{The Conditional Mass Function}
\label{sec:massfn}

In \paperone\ we derive $S(R)$ and $B(S)$ from simple theoretical considerations for all scales in a galactic disk. For a given turbulent power spectrum, with the assumption 
that the disk is marginally stable (Toomre $Q=1$), $S(R)$ is determined  
by summing the contribution from the velocity variance on all scales $R^{\prime}>R$
\begin{align}
\label{eqn:S.R}
S(R) &= \int_{0}^{\infty} 
|W(k,\,R)|^{2}
\ln{{\Bigl [}1 + \frac{3}{4}\,
\frac{v_{t}^{2}(k)}{c_{s}^{2} + \kappa^{2}\,k^{-2}}
{\Bigr]}} 
{\rm d}\ln{k} 
\end{align}
where $W$ is the window for the density smoothing,\footnote{For convenience 
we take this to be a $k$-space tophat inside $k<1/R$, which is implicit in our 
previous derivation, but we show in \paperone\ and \papertwo\ that this has little 
effect on our results.} $v_{t}(k)$ is the turbulent velocity dispersion averaged on a scale $k$ (trivially related to the turbulent power spectrum), $c_{s}$ is the thermal sound speed, and $\kappa$ is the epicyclic frequency ($=\sqrt{2}\,\Omega$, where $\Omega=V_{c}/R$ is the orbital frequency, for a disk with constant circular velocity $V_{c}$). $B(R)$ is 
\be
B(R) = \ln{\left(\frac{\rho_{\rm crit}}{\rho_{0}} \right)} + \frac{S(R)}{2}
\ee
where $\rho_{\rm crit}$ is the critical density above which a region is self-gravitating. This is
\begin{align}
\label{eqn:rhocrit}
\frac{\rho_{\rm crit}}{\rho_{0}} \equiv \frac{1}{2\,\tilde{\kappa}}\,\left(1+\frac{h}{R} \right)
{\Bigl[} \frac{\sigma_{g}^{2}(R)}{\sigma_{g}^{2}(h)}\,\frac{h}{R}  + 
\tilde{\kappa}^{2}\,\frac{R}{h}{\Bigr]} 
\end{align}
where $\rho_{0}$ is the mean midplane density of the disk, $h$ is the disk scale height, 
$\tilde{\kappa}=\kappa/\Omega=\sqrt{2}$ for a constant-$V_{c}$ disk, 
and 
\be
\sigma_{g}^{2}(R) = c_{s}^{2} + v_{\rm A}^{2} + \langle v_{t}^{2}(R) \rangle 
\ee 
($v_{\rm A}$ is the Alfv{\'e}n speed). The mapping between radius and mass is 
\be
M(R) \equiv 4\,\pi\,\rho_{\rm crit}\,h^{3}\,
{\Bigl[}\frac{R^{2}}{2\,h^{2}} + {\Bigl(}1+\frac{R}{h}{\Bigr)}\,\exp{{\Bigl(}-\frac{R}{h}{\Bigr)}}-1 {\Bigr]}
\ee
It is easy to see that on small scales, these scalings reduce to the Jeans criterion 
for a combination of thermal ($c_{s}$), turbulent ($v_{t}$), and magnetic ($v_{\rm A}$) 
support, with $M=(4\pi/3)\,\rho_{\rm crit}\,R^{3}$; on large scales it becomes the Toomre criterion 
with $M=\pi\Sigma_{\rm crit}\,R^{2}$.

Recall $\flast(S)\,{{\rm d}S}$ gives the differential fraction trajectories that have a last crossing in a narrow range ${\rm d}S$ about the scale $S[R]$ (corresponding to mass $M=M[R]$). Each trajectory randomly samples the Eulerian volume, so the differential number of last-crossing regions is related by $V_{\rm cl}(M)\,{\rm d}N(M) = V_{\rm tot}\,\flast(S[M])\,{{\rm d}S}$ (where $V_{\rm cl}(M) = M/\rho_{\rm crit}(M)$ is the cloud volume at the time of last-crossing and $V_{\rm tot}$ is the total volume sampled). Hence, the mass function -- the number density ${\rm d}n = {\rm d}N/V_{\rm tot}$ in a differential interval -- is given by
\be
\frac{{\rm d}n}{{\rm d}M} = 
\frac{\rho_{\rm crit}(M)}{M}\,\flast(M)\,{\Bigl |}\frac{{\rm d}S}{{\rm d}M} {\Bigr |}
\ee

Two parameters completely specify the model in dimensionless 
units. These are the spectral index $p$ of the turbulent velocity 
spectrum, $E(k)\propto k^{-p}$ (usually $p\approx5/3-2$), 
and its normalization, which we define by the Mach number on large scales 
$\mathcal{M}_{h}^{2}\equiv \langle v_{t}^{2}(h)\rangle/(c_{s}^{2}+v_{\rm A}^{2})$. 
The dimensional parameters $h$ (or $c_{s}$) and $\rho_{0}$ simply rescale 
the predictions to absolute units.

For a choice of $p$ and $\mathcal{M}_{h}$, it is straightforward to numerically 
determine $\flast(M\,|\,\delta_{0}[R_{0}])$, 
hence the conditional mass function (mass function within the appropriate sub-regions for any choice of $\delta_{0}$ and $R_{0}$). Unfortunately a closed-form solution is not generally possible. 
However, we show in \papertwo\ that on sufficiently small scales (near/below the 
sonic length $R_{\rm sonic} = h\,\mathcal{M}_{h}^{-2/(p-1)}$), 
the ``run'' in $S(R)\approx S(R_{\rm sonic})$ becomes small (since most of the power 
contributing in Eq.~\ref{eqn:S.R} comes from large scales), 
while $B(R)$ rises rapidly, so 
${\rm d}B/{\rm d}S \gg B(S)/S \gg 1$. In this limit, the conditional MF is 
approximately
\be
\label{eqn:approx}
\frac{{\rm d}n}{{\rm d}M}
\sim 
\frac{\rho_{\rm crit}(M)}{M^{2}}\,
{{\Bigl |}\frac{{\rm d}\ln{\rho_{\rm crit}}}{{\rm d}\ln{M}}{\Bigr |}}\,
P_{0}(\delta - \delta_{0}\,|\,S-S_{0}) 
%P_{0}(\ln{[\rho_{\rm crit}/\rho(S_{0})]}+(S-S_{0})/2\,|\,S-S_{0}) 
%\frac{1}{\sqrt{2\pi(S-S_{0})}}\exp{\left[-\frac{\Delta B^{2}}{2\,(S-S_{0})} \right]}
\ee
%\begin{align}
%\label{eqn:approx}
%\frac{{\rm d}n}{{\rm d}M}
%&\sim 
%\frac{\rho_{\rm crit}(M)}{M^{2}}\,
%{{\Bigl |}\frac{{\rm d}\ln{\rho_{\rm crit}}}{{\rm d}\ln{M}}{\Bigr |}}\frac{1}{\sqrt{2\pi(S-S_{0})}} \times\\
%\nonumber&\exp{\left[-\frac{(\ln{[\rho_{\rm crit}/\rho(S_{0})]}+(S-S_{0})/2)^{2}}{2\,(S-S_{0})} \right]}
%\end{align}
where
\begin{align}
P_{0}(\delta - \delta_{0}\, & |\,S-S_{0}) 
= \frac{1}{\sqrt{2\pi(S-S_{0})}} \times\\
\nonumber&\exp{\left[-\frac{(\ln{[\rho_{\rm crit}/\rho(S_{0})]}+(S-S_{0})/2)^{2}}{2\,(S-S_{0})} \right]}
%E_{0}(S-S_{0}) 
%= {\exp{\left[-\frac{(\ln{[\rho_{\rm crit}/\rho(S_{0})]}+(S-S_{0})/2)^{2}}{2\,(S-S_{0})} \right]}}
%\Delta B = \ln{\left[\frac{\rho_{\rm crit}}{\rho(S_{0})}\right]} + \frac{S-S_{0}}{2}
\end{align}

For the appropriate choice of $\delta_{0}$ on scale $R_{0}$, this is the resulting core MF (and so relates to the resulting stellar IMF and SFR) in over-dense regions (e.g.\ GMCs) or ``voids'' (inter-GMC gas). Taking $\delta_{0}\rightarrow0$ and $R_{0}\rightarrow\infty$, we recover the galaxy-averaged CMF. In a companion paper \citep{hopkins:excursion.imf.variation}, we consider in detail what this means for how the CMF (and by extension IMF) vary with environmental properties. For our purposes here, though, since we average over a range of masses, this has no effect on our results (and the predicted variation within a galaxy is very weak).

\begin{figure}
    \centering
    \plotonesize{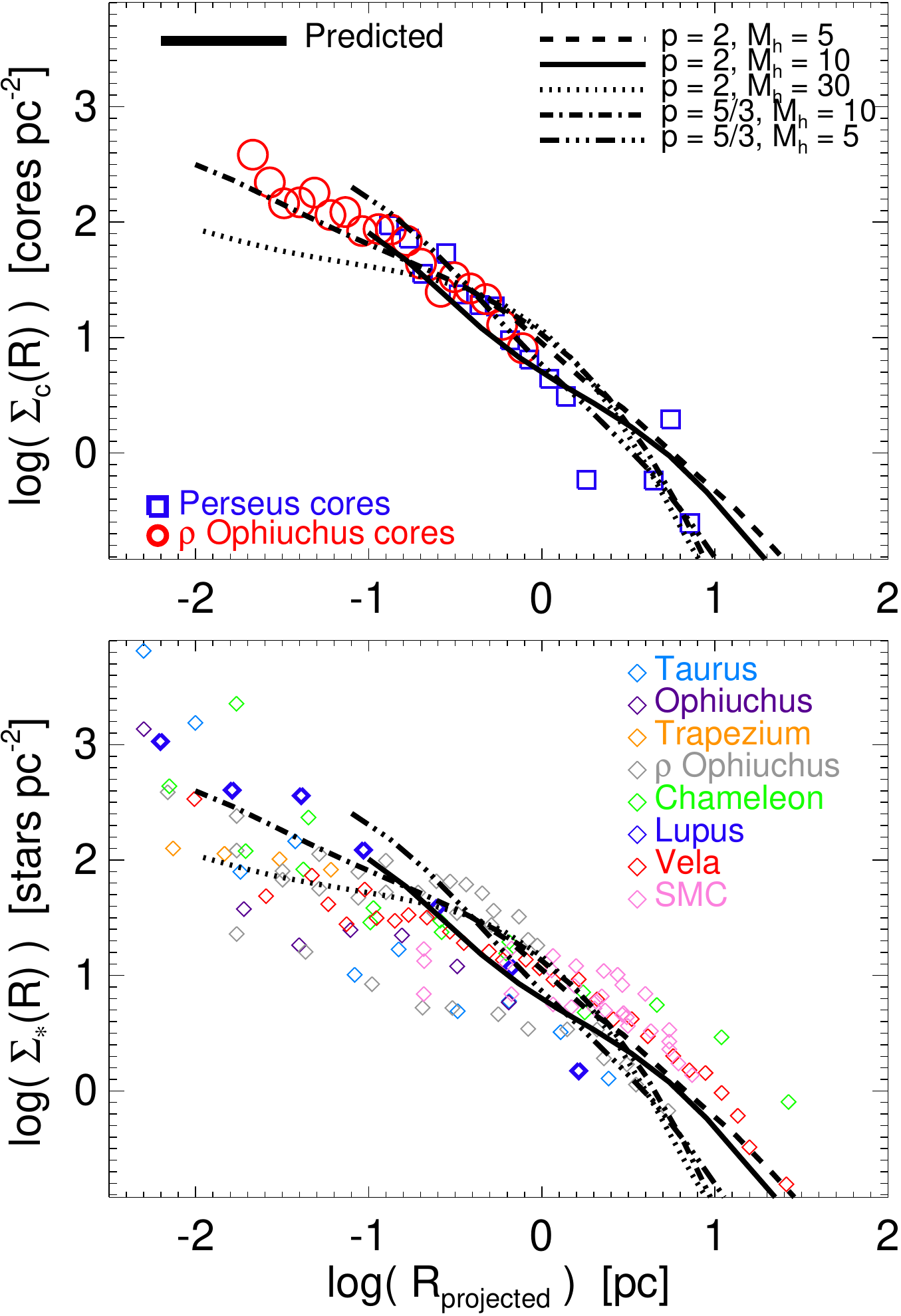}{1.0}
    \caption{Mean surface density $\Sigma_{\ast}(R)$ of stars around a random 
    star: this is proportional, by definition, to $1+\langle \xi_{\rm 2d}(R)\rangle$ 
    in Fig.~\ref{fig:corr.vs.param}. We compare the model predictions (linestyles as 
    Fig.~\ref{fig:corr.vs.param}; normalized by the mean surface density of the observed systems 
    and scale height $h=200\,$pc) to 
    observations of the corresponding core-core cross-correlation 
    ({\em top}; points from \citet{stanke:2006.core.mf.clustering,enoch:2008.core.mf.clustering}) and 
    a compilation of observations of the star-star cross-correlation 
    ({\em bottom}; points from \citet{simon:1997.stellar.clustering,nakajima:1998.stellar.clustering,hartmann:2002.taurus.stellar.clustering,hennekemper:2008.smc.stellar.clustering,kraus:2008.stellar.clustering}). The agreement is reasonably good over 
    the dynamic range observed, and the observational scatter in the shape of 
    $\xi_{\rm 2d}$ is similar to that predicted in Fig.~\ref{fig:corr.vs.param}. 
    The models do not extend to the sharp rise in stellar clustering 
    at scales $\lesssim 0.01$\,pc, usually attributed to binaries.
    \label{fig:crosscorr}}
\end{figure}

\vspace{-0.5cm}
\section{The Correlation Function}
\label{sec:corrfn}

The auto-correlation function $\xi$ of a given population is defined as the excess probability 
of finding another member of the population within a differential volume at a radius $r$ from one such member, 
i.e.\ 
\begin{align}
1 + \xi_{{\rm MM}}(r\,|\,M) \equiv \frac{\langle N(r\,|\,M) \rangle}{\langle n(M) \rangle {\rm d}V}
\end{align}
where $n(M) = {\rm d}n/{\rm d}M$ at mass $M$ and $N(r\,|\,M)$ is the differential number of objects in the mass range $M,\,M+{\rm d}M$ found at a radius $r$ from another object with mass $M$. But since $n(r\,|\,M)\propto \flast(M\,|\,\delta_{0}[S_{0}(r)])$, for a region with an 
overdensity $\delta_{0}$ on a scale $S_{0}$ (corresponding to $r$), 
this is just 
\be
1 + \xi_{{\rm MM}}(r\,|\,M) = \int_{\delta_{0}}\,{\Bigl (}\frac{\flast(M\,|\,\delta_{0})}{\flast(M)} {\Bigr)}\,
P_{\ast}(\delta_{0}\,|\,S_{0}[r])\,{\rm d}\delta_{0}
\ee
Here, $P_{\ast}(\delta_{0}\,|\,S_{0}[r])$ is the probability of $\delta(S_{0})$ having the value $\delta_{0}$ on the scale $S_{0}$, {\em given} that $\delta(S[M])=B(S[M])$ -- i.e.\ that there is a barrier crossing (a core/star) at the ``starting point.'' But the probability of having a last-crossing event at scale $S>S_{0}$, given a density $\delta_{0}(S_{0})$, is just $\flast(M\,|\,\delta_{0})$. So conversely, the probability of $\delta_{0}(S_{0})$ given 
that crossing is just related by Bayes's theorem
\be
P_{\ast}(\delta_{0}\,|\,S_{0}[r]) 
%= \frac{\flast(M\,|\,\delta_{0})\,P(\delta_{0})}{\langle P_{\rm lastcross}(M) \rangle} 
= \flast(M\,|\,\delta_{0})\,\frac{P_{0}(\delta_{0}\,|\,S_{0})}{\flast(M)}
\ee 
So we obtain 
\be
\label{eqn:autocorr}
1 + \xi_{{\rm MM}}(r\,|\,M) = \int_{-\infty}^{\infty}\,{\Bigl (}\frac{\flast(M\,|\,\delta_{0})}{\flast(M)} {\Bigr)}^{2}\,
P_{0}(\delta_{0}\,|\,S_{0}[r])\,{\rm d}\delta_{0}
\ee

More accurately, what is typically measured is the star-star or clump-clump correlation function over a broad mass range. We therefore require the cross-correlation between an initial crossing at $M_{1}$ and 
second crossing at some $M_{2}$. The logic is identical, however. We then integrate over the appropriate range of $M_{2}$ and, finally, average over $M_{1}$ (weighted by number density). 
We obtain 
\be
\label{eqn:xi.starstar}
1+\xi(r) = \frac{\int\,{\rm d}W_{M_{1}}\,
\int\,{\rm d}W_{M_{2}}\,
\int_{-\infty}^{\infty}\flast(M_{1}\,|\,\delta_{0})\,\flast(M_{2}\,|\,\delta_{0})\,
P(\delta_{0}\,|\,S_{0})\,
{\rm d}\delta_{0}
}
{\left(\int\,{\rm d}W_{M}\,\flast(M)\right)^{2}} 
\ee
where 
\be
{\rm d}W_{M} \equiv \frac{\rho_{\rm crit}(M)}{M}\,{\Bigl |}\frac{{\rm d}S(M)}{{\rm d}M}{\Bigr |}\,{\rm d}M
\ee
and the integrals over mass $M$ should be over the appropriate observed core/stellar mass range.

Finally, note that what is generally measured is the projected correlation function $\xi_{2d}(R_{p})$. 
Projecting $\xi(r)$ is straightforward:
\be
\label{eqn:xi.2d}
\xi_{2d}(R_{p}) = 
\frac{\int_{-\infty}^{\infty}\,n_{0}(z)\,\xi_{3d}(\sqrt{R_{p}^{2}+z^{2}})\,{\rm d}z}
{\int_{-\infty}^{\infty}n_{0}(z)\,{\rm d}z}
\ee
where $z$ is the line-of-sight direction and $n_{0}(z)$ is the 
average abundance expected. Technically this will depend on projection angles 
(especially in the complicated case of systems inside the MW), but for almost any profile 
where $n_{0}\approx n(M)$ is a weak function of $z$ and then falls off 
at scales $\gtrsim h$ the results are nearly identical.

\vspace{-0.5cm}
\section{Results}
\label{sec:results}

In Fig.~\ref{fig:autocorr}, we plot the exact (3d) autocorrelation function 
from Eq.~\ref{eqn:autocorr} as a function of 
$r$ and $M$. Recall, in dimensionless units the problem is completely specified by a 
choice of $p$ and $\mathcal{M}_{h}$, for which we adopt canonical 
values $p=2$ (Burgers turbulence, typical in the supersonic regime) and 
$\mathcal{M}_{h}=10$ (typical of the cold clumps in the MW disk). 

The exact results in Fig.~\ref{fig:autocorr} require the numerical solution for 
$\flast(M\,|\,\delta_{0})$. 
However, if we restrict to the regime where 
${\rm d}B/{\rm d}S\gg B/S \gg 1$ so that the mass function 
can be approximated by Eq.~\ref{eqn:approx}, or consider a linear barrier (Eq.~\ref{eqn:linear}), 
then this has the closed-form solution
\begin{align}
\label{eqn:autocorr.approx}
%1 + \xi(r\,|\,M) &\approx \frac{1}{\sqrt{1-\tilde{s}^{2}}}\,
%\exp{\left(\frac{B^{2}}{S\,(1+\tilde{s}^{-1})} \right)}\\
%\tilde{s} &\equiv \frac{S_{0}}{S}\,,\ \ \ \ \ \ \ ({\rm d}B/{\rm d}S\gg1)
1 + \xi_{\rm MM}(r\,|\,M) &\approx \frac{1}{\sqrt{1-(S_{0}/S)^{2}}}\,
\exp{\left(\frac{B^{2}}{S\,(1+S/S_{0})} \right)}
\end{align}
At small and large $r$, respectively, this becomes
\begin{align}
\label{eqn:limit.small}
\xi_{\rm MM}(r\,|\,M) &\rightarrow \frac{\exp{(B^{2}/2S)}}{\sqrt{2\,(1-S_{0}/S)}}\ \ \ \ \ (S_{0}\rightarrow S) \\
\label{eqn:limit.large}
\xi_{\rm MM}(r\,|\,M) &\rightarrow {\Bigl (} \frac{B}{S} {\Bigr)}^{2}\,S_{0}\ \ \ \ \ (S_{0}\rightarrow 0)
\end{align}
We compare this to the exact result. The shape of $\xi(R)$ and its systematic dependence 
on $R$ are qualitatively described. However, it is nowhere exact, because $S(R)$ does vary significantly on larger scales, and $B(R)$ and $S(R)$ both vary such that even if ${\rm d}B/{\rm d}S\gg1$, ${\rm d}B/{\rm d}S\gg B/S$ is less true.

The shape of $\xi_{\rm MM}(r)$ is similar in all cases -- i.e.\ it is nearly mass-independent over the interesting range -- so even though the expression for the CMF-averaged cross-correlation $\xi(r)$ in Eq.~\ref{eqn:xi.starstar} is complicated, it mostly amounts to a MF-weighted normalization. What we care about is the correlation function shape (we note below that the observed normalization is arbitrary and should be factored out in any case). This means that regardless of the CMF shape used to ``average'' the prediction, the result is similar. For the same reason, provided the general assumption that stars form from self-gravitating cores is reasonable, there will be no large differences between the predicted stellar and core correlation functions, regardless of how this process occurs. If the conversion from core mass to stellar mass were constant, the correlation functions would be exactly identical; but even allowing for a large random variation in conversion efficiency or systematic mass dependence makes little or no difference to our conclusions. We have, for example, considered sampling Eq.~\ref{eqn:xi.starstar} over only different sub-ranges in mass (say sampling only massive stars), or sampling over the full stellar IMF but with a random $0.5\,$dex scatter in the assumed stellar-to-core mass ratio, or allowing for the stellar-to-core mass ratio to depend on mass $\propto M^{-1}$; in all cases the differences in the predicted $\xi(r)$ shape are smaller than the differences owing to different choices of the turbulent spectrum. What will change the clustering, on very small scales $\ll R_{\rm sonic}$, is if individual single Jeans-mass cores form multiple stars. And to some extent, this must happen, as many stars are in binary systems. This is not accounted for here; therefore the predictions cannot be reliably extended to scales below the characteristic core scale $\lesssim R_{\rm sonic}$ (and indeed we see a discrepancy appear at these scales). We discuss this further below.

In analogy to the clustering of dark matter halos and GMCs defined in 
\paperone, we can define the linear bias as the ratio of the autocorrelation to 
the autocorrelation function of the mass itself. On large scales, this should approach 
a constant, 
\be
b(M)^{2} \equiv \xi / \xi_{\rm mass}\ \ \ \ \ (r\rightarrow\infty)
\ee
On large scales (where $S_{0}[r]$ is small), the autocorrelation of the mass is just 
the variance $\xi_{\rm mass} \approx S_{0}$ 
\citep[see][]{mowhite:bias}. 
From Eq.~\ref{eqn:limit.large}, we can therefore immediately obtain the bias 
in the limit 
${\rm d}B/{\rm d}S\gg B/S \gg 1$, 
\be
b(M)\approx\frac{B(S)}{S}\sim 1 - 
   \left\{ \begin{array}{ll}
      0.04\,\ln{(M/M_{\rm sonic})} & M \gtrsim M_{\rm sonic} \\
      0.17\,\ln{(M/M_{\rm sonic})} & M \lesssim M_{\rm sonic}\ 
\end{array}
    \right.
\ee
where the latter equality is for typical parameters ($p=2$, $\mathcal{M}_{h}\approx30$) and $M_{\rm sonic}=M(R_{\rm sonic})$. The large-scale bias is a very weak function of mass and about unity.

In Fig.~\ref{fig:corr.vs.param} we plot the projected 
$\xi_{\rm 2d}$ 
from Eq.~\ref{eqn:xi.2d}, for the full CMF/IMF-averaged $\xi(r)$.\footnote{Technically, we 
integrate the CMF range from $10^{-4}-10\,M_{\rm sonic}$,  
which for typical parameters corresponds to $\sim10^{-3}-100\,\msun$, but we stress that so long as we 
include the ``peak'' of the CMF/IMF, this choice makes little difference except in the normalization of $\xi_{\rm 2d}$. To project, we also assume the 
average abundance $n_{0}(z)$ is flat out to $\pm2\,h$, but changing this limit or assuming 
an exponential $n_{0}\propto\exp(-z/h)$ has only weak effects.}
We examine the effects of the two free model parameters: the large-scale Mach number 
$\mathcal{M}_{h}$ and turbulent spectral index $p$. As shown in \papertwo\ for 
the CMF/IMF, changing $p=5/3$ instead is nearly equivalent to assuming a higher $\mathcal{M}_{h}$ 
at $p=2$ (since, in both cases, the sonic length is pushed down and turnover at low 
masses is slower). At lower $\mathcal{M}_{h}$, decreasing $\mathcal{M}_{h}$ increases 
the clustering signal, because there is a smaller variance $S$, so the small-scale 
cores depend more heavily on large-scale fluctuations. At sufficiently large $\mathcal{M}_{h}$, 
however, there is a significant change in shape in $\xi_{\rm 2d}$; the initial rise below $\sim h$ 
is rapid, but near $R_{\rm sonic}$, $\xi_{\rm 2d}$ flattens. This is a projection effect: the rise 
in $\xi_{\rm 3d}$ on small scales is still similar (just slightly more shallow) to that we see in $\xi_{\rm 2d}$ at smaller $\mathcal{M}_{h}$; however there is much more power on intermediate scales as well, so the 
power in $\xi_{\rm 2d}$ is nearly ``converged,'' hence the flattening. 

In Fig.~\ref{fig:crosscorr} we compare the projected $\xi_{\rm 2d}$ to a compilation of observations of both the star-star and core-core autocorrelation functions. Typically, the observations are 
plotted not as $\xi_{\rm 2d}$ but as 
$\Sigma_{\ast}$, defined as the average surface number density ${\rm dN}/{\rm dA}(R_{p})$ of 
companions at a radius $R_{p}$ from the primary. By definition, this is trivially related to 
$\xi_{\rm 2d}$ as 
\be
\Sigma_{\ast}(R_{p}) = \langle\Sigma_{\ast}\rangle(1 + \xi_{\rm 2d})
\ee
Here, $\langle\Sigma_{\ast}\rangle$ is undetermined -- it depends both on the absolute background galaxy properties which set the scale of the problem ($h$, $\rho_{0}$, etc) and on the uncertain star formation efficiencies (fraction of gas which will actually turn into stars). We treat it as arbitrary and simply compare the profile shape. We normalize $R$ to an absolute physical scale by assuming a MW-like scale height $h=200\,$pc for the  gas.

We can gain some (approximate) insight into the functional form of $\xi_{\rm 2d}$ on small scales from Eq.~\ref{eqn:autocorr.approx}. Assume $r\ll h$ (the dynamic range of most interest), in the limit of Eq.~\ref{eqn:limit.small}, so $\mathcal{M}^{2}(r)\ll \mathcal{M}_{h}^{2}$ as well, and expand to leading order in $\mathcal{O}(r/h)$ and $\mathcal{O}(\mathcal{M}_{h}^{-1})$ (series expanding Eq.~\ref{eqn:S.R} around $S=S_{0}$); finally make the simple approximation that projection to $\xi_{\rm 2d}$ multiplies $\xi(r)$ by one power of $R$ and that we can treat the mass-averaging as a normalization correction. After some tedious algebra, we obtain 
\begin{align}
1+\xi_{\rm 2d} &\propto \frac{(1+\frac{3}{8}\,x^{p-1})^{1/4}}{x^{(p-1)/2-1}}\,
{\Bigl [}\frac{\mathcal{M}_{h}^{-2+\frac{4}{(p-1)}}}{\sqrt{2}}
\frac{(1+x^{p-1})}{x^{2}}
{\Bigr ]}^{
\frac{1}{p-1} + \epsilon
}\\
&\propto \frac{(1+x)\,(1+\frac{3}{8}\,x)^{1/4}}{x^{3/2}} + \mathcal{O}(\ln^{-1}{\mathcal{M}_{h}})\ \ \ \ \ \ \ \ (p=2) 
%&\propto \frac{(1+x^{2/3})^{3/2}\,(1+\frac{3}{8}\,x^{2/3})^{1/4}}{x^{7/3}} + \mathcal{O}(\ln^{-1}{\mathcal{M}_{h}})\ \ \ \ \ (p=5/3) \\ 
%\alpha &\equiv \frac{1}{4\,\ln{\mathcal{M}}}\,\ln{(x^{-2}\,(1+x^{p-1}))} \\ 
%x &\equiv \frac{R}{R_{\rm sonic}} = \frac{R}{h\,\mathcal{M}_{h}^{-2/(p-1)}}
\end{align}
where $x \equiv R/R_{\rm sonic}$, and $\epsilon = \ln{(x^{-2}\,(1+x^{p-1}))}/4\,\ln{(\mathcal{M}_{h})}$. Although we have made a number of simplifications, we recover all the key behaviors in Fig.~\ref{fig:corr.vs.param}: the predicted $\xi_{\rm 2d}(R)$ is an approximate power-law (because $\rho_{\rm crit}(r)$ is so), which scales steeply ($\propto R^{-(1.5-2.3)}$ for $p=5/3-2$) on small scales, then flattens around the sonic length (first to $\propto R^{-(0.5-1.0)}$ then $R^{-0.25}$). The ``steepening'' below $R_{\rm sonic}$ is caused by the steeper dependence of $\rho_{\rm crit}$ on $R$ at these scales (thermal support preventing collapse); the gradual run at larger scales from the logarithmic run in $S$ with $\mathcal{M}(r)$. 

Although it is less precisely defined, we can also make some estimate of the fraction of stars formed in an ``isolated'' (non-clustered) mode. Specifically, in analogy to our derivation of Eq.~\ref{eqn:autocorr}, we begin with the probability that a core (last-crossing event) at $\flast(S[M])$ is embedded in a larger region $S_{0}[r]$ with some $\delta_{0}(S_{0})$; we can then calculate the probability that this region contains zero additional last-crossing events. This latter probability is just given by $1-\int_{S_{0}}^{S_{i}}{\rm d}\Sprime\,\flast(\Sprime\,|\,\delta_{0})$ from Eq.~\ref{eqn:integral.eqn}, i.e.\ one minus the total probability of a crossing at $S>S_{0}$. After some algebra, we obtain the probability of a core of mass $M$ having no additional crossings within the parent radius $r[S]$: 
\begin{align}
f_{\rm iso}(<r\,|\,M) = 1 - \int_{-\infty}^{\infty}{\rm d}\delta_{0}\,\int_{S_{0}}^{S_{i}}{\rm d}\Sprime\,
\frac{\flast(S\,|\,\delta_{0})}{\flast(S)}\,
\flast(\Sprime\,|\,\delta_{0})\,P_{0}(\delta_{0}\,|\,S_{0}[r])
\end{align}
If we take the limits ${\rm d}B/{\rm d}S\gg B/S \gg 1$, we can obtain the approximate scaling $f_{\rm iso}(r<r_{1}\,|\,M[r_{0}]) \sim \ln^{-1}(r_{1}/r_{0})\,(r_{1}/r_{0})^{-{\rm d}B/{\rm d}S}$. In greater detail, if we solve this for a linear barrier and and assume the less restrictive ${\rm d}B/{\rm d}S\gtrsim1$ and $r\ll h$, and insert $B(r)$ and $S(r)$ defined above, we finally obtain
\begin{align}
f_{\rm iso}(<r\,|\,S) &\approx \frac{2}{3}\,{\Bigl |}\frac{{\rm d}B}{{\rm d}S}{\Bigr |}^{-1}\,P_{0}(B[S]-B[S_{0}]\,|\,S-S_{0}[r]) \\
% + \mathcal{O}{\Bigl(} {\Bigl |}\frac{{\rm d}B}{{\rm d}S}{\Bigr |}^{-2} {\Bigr)} \\
&\sim 0.15\,{\Bigl[}\ln{\Bigl(1+\frac{r}{R_{\rm sonic}} \Bigr)}{\Bigr]}^{-1/2}\,
{\Bigl(} \frac{r}{R_{\rm sonic}}{\Bigr)}^{-0.65}
\end{align}
where in the latter we insert $p=2$ and $\mathcal{M}_{h}\sim5-30$ (the result is very weakly dependent on $\mathcal{M}_{h}$). This should translate to the fraction of cores formed without a companion core inside of $r$. We caution that if only a small fraction of cores make stars, the fraction of stars formed without another star inside $<r$ will be higher. But if cores produce multiple stars (recall that we do not explicitly treat binaries here), the fraction of isolated stars will be even lower. Regardless of these uncertainties, we see that the absolute number of cores/stars formed without neighbors inside a radius $r$ is small even for $r$ near the sonic length, and falls rapidly as we expand the ``search radius.''

\vspace{-0.5cm}
\section{Discussion}
\label{sec:discussion}

We have developed an analytic theory for the clustering properties of protostellar cores in a supersonically turbulent density field.
In \paperone, we developed an excursion-set theory for lognormal density fluctuations in the ISM, 
and applied it to the mass functions and clustering of GMCs -- the ``first-crossing distribution'' (distribution of bound masses on the largest self-gravitating scales). In \papertwo, we showed that this could be extended to predict the protostellar core MF by defining the ``last-crossing distribution''  (distribution of masses on the {\em smallest} scales on which they are self-gravitating and non-fragmenting). 
Here, we extend this method to define the conditional last-crossing mass function, as a function of density on larger scales: i.e.\ the ``two-barrier last crossing problem.'' We use this to analytically derive the correlation function of these cores as a function of separation (and core mass), on all scales from the core itself to galactic scales. This should directly translate to the correlation properties of newly-formed stars, independent of the mass conversion/star formation efficiency. 

We find that, on {\em large} scales $\gtrsim h$ (the disk scale height), cores are weakly biased. Core formation, therefore, is less sensitive to galaxy-scale overdensities, except insofar as those overdensities collect/drive the dense gas. This is related to our derivation in \paperone\ of the clustering of GMCs: they themselves are not strongly biased (relative to the rest of the dense gas) on large scales, provided there is sufficient gas present and cooling is efficient. But we caution that on scales $\gtrsim h$, global modes such as spiral waves may be important, and we cannot assume quantities like $\Sigma$ or $\Omega$ are constant. So the relation to e.g.\ the global Schmidt-Kennicutt law and observations of SF efficiency versus galaxy structure \citep[e.g.][]{leroy:2008.sfe.vs.gal.prop,foyle:2010.spiral.arms.conc.gas.not.sf} is only approximate. 

On scales $\ll h$, however, cores cluster very strongly. The formal three-dimensional clustering amplitudes at small scales are enormous -- in other words, stars form in a clustered manner. The approximate conditional mass function (on small scales) in Eq.~\ref{eqn:approx} illustrates why. On small scales, the ``run'' in $S(R)$ given by Eq.~\ref{eqn:S.R} (hence $\Delta S\equiv S-S_{0}$ in Eq.~\ref{eqn:approx}) is small, because $\mathcal{M}^{2}(R)$ is small. As this vanishes, the $P_{0}$ term in Eq.~\ref{eqn:approx} becomes extremely sharply peaked around $\rho(R_{0})=\rho_{\rm crit}(R)$ -- so the CMF/IMF-averaged conditional mass function approaches a step function in $\rho(R_{0})$ on some larger scale. Given the same arguments, the fraction of stars formed in an ``isolated'' mode should scale as $\mathcal{O}({\rm d}B/{\rm d}S)^{-1}$; near the sonic length this is approximately $|{\rm d}B/{\rm d}S|_{R=R_{\rm sonic}}^{-1}\lesssim 0.1$.

Physically, the meaning of this is: as we average on smaller scales approaching the sonic length, the local rms Mach numbers decline rapidly ($\mathcal{M}^{2}\propto r^{p-1}$), so the contribution to density fluctuations (the variance $S(R)$, which is directly tied to the Mach number in supersonic turbulence) also drops. But the threshold density for collapse continues to rise. So the smallest scale at which a region is going to be self-gravitating (or avoid fragmentation) is essentially ``imprinted'' by the density by fluctuations driven on larger scales. It is unlikely to ``wander'' much across the barrier on smaller scales since fluctuations are small. 

As a result, stars preferentially form in larger-scale overdense regions which must themselves (on some scale) be self-gravitating. The characteristic scale of the ``parent'' systems -- i.e.\ the characteristic scale at which the clustering amplitude will fall off, is the scale where $S(R)$ begins to run significantly with $R$. But this is by definition the characteristic scale of the {\em first-crossing} distribution, which we identified in \paperone\ with GMCs, and showed there has a characteristic scale of $\sim h$ -- just set by the global turbulent Jeans length. In short, stars form inside of GMCs, in strongly clustered fashion, with the characteristic length of that clustering set by $\sim h$. 

These arguments are quite general: the fundamental statement is that, since most of the power in the turbulent velocity field is on large scales ($\sim h$), star formation must be strongly clustered below that scale. This is true for any plausible turbulent power spectrum. We show explicitly that a qualitatively similar scaling results for turbulent spectral indices $p\sim5/3-2$, and large-scale Mach numbers $\mathcal{M}_{h}\sim5-30$. 
However the details of the correlation function do depend on these values, in a non-trivial manner. 
For example, the slope on small scales $\xi_{\rm 2d}\propto R_{p}^{-\alpha}$, is an approximate power-law with values $\alpha\approx0.5-1.0$; it can flatten significantly to $\alpha\approx 0.1-0.5$ on sufficiently small scales if the Mach numbers are sufficiently large.\footnote{If power were not concentrated on large scales, for example if the variance $S$ rose steeply and continuously down to small scales, then at almost all scales around a core/star we would have $S_{0}\ll S$ in Eq.~\ref{eqn:autocorr.approx}, hence predict $1+\xi \approx 1 + (B^{2}/S)\,(S_{0}/S) = 1+\nu^{2}\,(S_{0}/S)$, where $\nu$ represents the number of standard deviations needed for a core to form. Thus, if cores can ``easily'' form near the mean density ($\nu\lesssim1$), and/or the power in the density field continued to rise down to small scales ($S_{0}\ll S$), $\xi\ll1$ and $\Sigma_{\ast}(R_{p}) = \langle \Sigma_{\ast}\rangle(1+\xi) \approx \langle \Sigma_{\ast}\rangle$ would be nearly flat as a function of radius, in stark disagreement with the observed clustering. For example, for $p=1.1$ and a Toomre $Q=0.01$, $\Sigma_{\ast}(R_{p})$ in Fig.~\ref{fig:crosscorr} increases by a factor $<2$ from $R=500$\,pc to $R=0.01$\,pc.}

We compare the predicted correlation functions to observations of both protostellar cores and young stars, and find that they agree well on the applicable scales. The scatter in the shape of observed correlation functions is also intriguingly similar to the range predicted from the parameter variations above.\footnote{It does appear that in some systems, the stellar correlation function may be slightly flatter on small scales than the core correlation function, but this will inevitably occur in time as stars move from their birthplace, and so it is difficult to disentangle without knowing the velocity and age distributions of the stars.}

On the smallest scales $\ll0.1\,$pc, our prediction does not rise as steeply as the observed stellar correlation functions. This is expected, since this effect comes primarily from binary and multiple stellar systems, which we do not explicitly predict since the fragmentation events are on a sub-core scale. Our derivation implicitly samples a ``snapshot'' within fully developed turbulence. We would need to treat cores collapsing in time to follow successive fragmentation and/or stellar accretion as the cores contract and form stars, but this will in turn depend in detail on the gas thermodynamics and stellar feedback \citep[see e.g.][]{krumholz:2009.massive.sf.accretion,peters:2010.massive.sf.accretion}. In \papertwo, we discuss in detail how this (and other time-dependent processes) may modify the relation between CMF and IMF; if the effect on the density distribution and/or multiplicity is truly scale-free, these effects factor out in our analysis here, but there is no strong reason to believe it would be so. However, on larger scales, theoretical arguments (and some observations) suggest that the primary role of feedback is to regulate the turbulent cascade and generate outflows, as well as to regulate the core-to-stellar mass conversion efficiencies \citep[see e.g.][]{matzner:2000.lowmass.sf.eff,mac-low:2004.turb.sf.review,veltchev:2011.frag,alves:2007.core.mf.clustering,enoch:2008.core.mf.clustering}; in that regime, our conclusions should be robust. Improving the predictions here by taking these processes into account may be possible in the time-dependent formulation of the excursion set ISM model developed in \paperone, since that could follow the simultaneous growth of fluctuations and collapse of a given sub-region, coupled to appropriate models for the thermodynamics and stellar feedback.

\vspace{-0.7cm}
\acknowledgments 
We thank Chris McKee and Eliot Quataert for helpful discussions in the development of this work, as well as Eli Bressert, Stella Offner, and our referee, Federico Pelupessy, for a number of suggestions and comments. Support for PFH was provided by NASA through Einstein Postdoctoral Fellowship Award Number PF1-120083 issued by the Chandra X-ray Observatory Center, which is operated by the Smithsonian Astrophysical Observatory for and on behalf of the NASA under contract NAS8-03060.
\\

\vspace{-0.1cm}
\bibliography{/Users/phopkins/Documents/lars_galaxies/papers/ms}

\end{document}